\documentclass[pre,aps,twocolumn,showpacs,superscriptaddress]{revtex4}
\usepackage{graphicx}
\begin{document}

\title{Packing geometry and statistics of force networks
in granular media}

\author{Jacco H. Snoeijer} \affiliation{Instituut-Lorentz,
Universiteit Leiden, Postbus 9506, 2300 RA Leiden, The
Netherlands}

\author{Martin van Hecke} \affiliation{Kamerlingh Onnes Lab,
Universiteit Leiden, Postbus 9504, 2300 RA Leiden, The
Netherlands}

\author{Ell\'ak Somfai}
\affiliation{Instituut-Lorentz,
Universiteit Leiden, Postbus 9506, 2300 RA Leiden, The
Netherlands}

\author{Wim van Saarloos} \affiliation{Instituut-Lorentz,
Universiteit Leiden, Postbus 9506, 2300 RA Leiden, The
Netherlands}

\date{\today}
\begin{abstract}
The relation between packing geometry and force network statistics 
is studied for granular media. 
Based on simulations of two-dimensional packings of 
Hertzian spheres, we develop a geometrical framework relating 
the distribution of interparticle forces $P(f)$ to the weight 
distribution ${\cal P}(w)$, which is measured in experiments. 
We apply this framework to reinterpret recent experimental data 
on strongly deformed packings, and suggest that 
the observed changes of ${\cal P}(w)$ are dominated by changes 
in contact network while $P(f)$ remains relatively unaltered. 
We furthermore investigate the role of packing disorder in the 
context of the $q$-model, and address the question of how 
force fluctuations build up as a function of the distance 
beneath the top surface.
\end{abstract}


\maketitle
\section{Introduction}\label{sec.introduction}
Inside a granular material forces are distributed very
inhomogeneously: a small number of particles carries a large
fraction of the internal forces \cite{gm}. These large
fluctuations are reflected in the force probability density
functions, which typically decay exponentially 
\cite{network,exp2,Pf,qm}. 
The behavior for small forces is not as well
understood as the generic exponential tail: the $q$-model
appears to predict a vanishing probability density for small
forces \cite{qm}, whereas experiments and simulations clearly show that
this probability remains non-zero \cite{network,exp2,Pf}. 
The characterization and understanding of this probability 
remains a challenge, especially since the force distribution 
is believed to play an important role for the dynamical arrest 
or ``jamming'' of granular and other disordered materials \cite{liuletter}. 
In particular, the force distribution has been observed to develop a small
peak (around the average value) in simulations of supercooled
liquids, foams and granular matter undergoing a jamming
transition \cite{liuletter,grestjam}. 
However, there is still no microscopic understanding how this effect 
relates to the properties of the force network.

\begin{figure}[t]
\includegraphics[width=8.5cm]{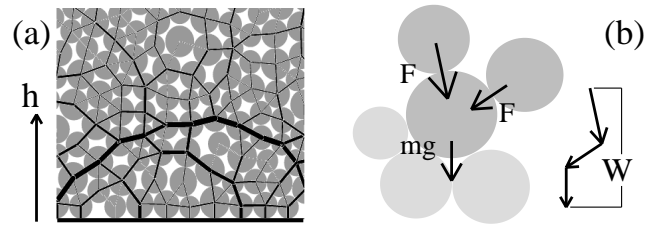}
\caption{{\em (a)} Detail of
 a typical packing in our simulations; the height $h$ denotes the
 distance from the bottom. The force network is represented by the
 black lines whose thickness is proportional to the force-magnitude. {\em (b)}
 Definition of interparticle forces $F$ and weight $W$, for a 
 frictionless particle with $n_c \!=\!2$.}
\label{fig.weights}
\end{figure}

This paper is a full exposition and expansion of a new approach, 
which was briefly outlined in \cite{heggn}. 
We will unravel the effect of the local contact geometry
on the distributions of {\em interparticle force} $F $ and effective
particle {\em weight} $W $; the weight is defined as the sum of the
vertical components of all downward pointing forces on a particle -- 
see Fig.~1. 
While the distribution of forces $F$ is the primary object one 
ultimately wishes to characterize, it is difficult to access 
experimentally. 
Experiments with photoelastic materials are able to depict the 
spatial structure of bulk forces in 2D, but their precision 
to resolve individual contact forces is limited \cite{photoel}. 
Only recently, there have been first reports of 3D bulk measurements 
on forces in compressed emulsions \cite{edwards3D}. 
Most quantitative information on the force probability distribution 
is at present only accessible through measurements 
of the {\em particle-wall forces} from imprints on carbon 
paper \cite{network} or by force sensors \cite{exp2}. 
Each particle-wall force has to balance all 
interparticle forces that are exerted on the corresponding particle
from above, see Fig.~\ref{fig.weights}. This means that experiments
essentially measure a combination of forces that we refer to as the
{\em weights} of the bottom particles. For simplicitly, we will focus 
on frictionless spheres for which these weights are defined as
\begin{equation}\label{defweight}
W_j \equiv m_j g + \sum_{<i>} (\vec{F}_{ij})_z~. \label{defw}
\end{equation}
Here $m_j$ denotes mass, $g$ denotes gravity, $\vec{F}_{ij}$ are the
interparticle forces and $n_c$ is the number of particle exerting a
force on particle $j$ {\em from above}; the sum runs over all these
forces. So, to relate the experimental results to the bulk force
distributions, one has to understand the relation between weights and
forces. 

In this paper we will show how the {\em local packing geometry} plays
the crucial role in the relation between the force distributions
$P(f)$ and the weight distributions ${\cal P}(w)$ (we define
$f=F/\langle F\rangle$ and $w=W/\langle W\rangle$ as the 
appropriately rescaled forces and weights). Our central point
is that while the distribution of $f$ is robust, the distribution of
$w$ is profoundly influenced by the contact geometry, in particular by
{\em the number of downward pointing contact forces} $n_c$. In simulations of 
Hertzian sphere packings we will find that 
${\cal P}_{\it boundary}(w)$ is different from ${\cal P}_{\it bulk}(w)$, 
due to the rather special packing geometry near a boundary. 
However, for many (but not all) experimentally relevant
situations, the special packing geometry near a boundary makes 
${\cal P}_{boundary}(w)$ rather close, but not equal, to the bulk
$P(f)$. This fortunate but non-trivial coincidence can be
understood easily within our framework. We will, however, also provide 
two examples
where ${\cal P}_{\it boundary}(w)$ and bulk $P(f)$ are significantly
different.

Additional motivation for studying the relation between forces,
weights and geometry comes from the $q$-model \cite{qm}. 
Once the distinction 
between forces and weights has been made, one notices that the 
$q$-model is a lattice model in which {\em weights} are randomly 
redistributed over a fixed number of supporting grains. 
The $q$-model displays a weight
distribution that is qualitatively different from both experimentally
observed weight distributions, or numerically obtained force
distributions. We will show that this is due to the fixed
connectedness of the $q$-model. Realistic ${\cal P}(w)$ can be obtained 
if we allow for the connectivity to vary within the $q$-model, e.g. 
by introducing random connectivity.

Our work then serves three purposes. 
First of all, it helps to interpret 
data obtained by measurements of particle-wall forces: 
this paper includes a section where we explicitly apply our framework 
to recent experimental data of highly compressed packings \cite{rubber}. 
Secondly, it shows how the simple $q$-model can be extended to obtain 
very realistic weight distributions for both regular and irregular 
packings. Since the model is known to give incorrect predictions on 
spatial propagation \cite{greens}, our intention is not to fine-tune 
the model and its parameters, but rather to indicate how the contact 
geometry is essential to describe force and weight fluctuations in 
more realistic packings. 
Thirdly, we address the question of how force fluctuations build up 
as a function of the distance beneath the top surface, providing 
another fundamental test for theoretical models. 

The paper is organized as follows. 
In Sec.~\ref{sec.pf} we first explain our numerical
method and then discuss the force distributions observed in amorphous 
packings: it turns out that $P(f)$ is rather insensitive to the packing 
geometry. 
We then show in Sec.~\ref{sec.pw} that the weight distributions 
${\cal P}(w)$, on the other hand, are very sensitive to the packing 
geometry. Using simple phase space considerations, 
we relate ${\cal P}(w)$ to $P(f)$ for a given geometry. 
This provides a recipe how to reconstruct the bulk $P(f)$ from the 
experimental data, and in Sec.~\ref{sec.exp} we explicitly apply this 
to recent experimental data on highly compressed packings \cite{rubber}. 
In particular, our analysis strongly suggests that $P(f)$ is essentially 
unaffected by the tremendous deformations encountered in the experiments. 
We then indicate some limitations of our framework in Sec.~\ref{beyond}, 
where we address subtle packing problems like the effect of gravity. 
In Sec.~\ref{sec.qmodel1} we investigate to what extent the
$q$-model can describe the results of the numerical packings 
of Hertzian spheres: 
we derive a surprising exact result for the bond quantities $qw$, 
and we investigate the role of disorder in the packing geometry. 
Finally, we address the top-down relaxation of force fluctuations 
in Sec.~\ref{sec.qmodel2}. We find no evidence in the Hertzian sphere 
packings for the power-law relaxation predicted by the $q$-model, 
indicating that the model is not able to capture this spatial aspect 
of the force network. The paper ends with a discussion.

\section{Statistics of interparticle forces}\label{sec.pf}

In this section we study the distribution of interparticle forces via
simulations of 2D packings of frictionless spheres. After introducing
our numerical method in Sec.~\ref{subsec.method}, we discuss the
similarities between $P(f)$ in the bulk and near the boundary 
(Sec.~\ref{subsec.pf}). 
We also study the angular distribution and the probability
distribution of the $z$ components of the contact forces in 
Sec.~\ref{subsec.orientation}, and close with a brief summary of results 
in section \ref{subsec.pfconclusion}.

\subsection{Numerical method and parameters}\label{subsec.method}
Our two-dimensional packings consist of frictionless spheres (3D) 
under gravity. 
The packings are created from molecular dynamics simulations
of spheres that interact through normal Hertzian forces, where
$F\propto \delta^{3/2}$ and $\delta$ denotes the overlap distance 
\cite{hertz}. Since Hertz's law for 2D disks is linear in $\delta$, 
we use 3D spheres. These particles reside 
in a container that is $24$ particle diameters wide, 
with periodic boundary conditions in the horizontal direction. 
The bottom support is rigid and also has a frictionless Hertzian 
interaction with the particles. We construct our stationary packings
by letting the particles relax from a gas-like state by
introducing a dissipative force that acts whenever the overlap
distance is non-zero. In this paper we use two different
polydispersities: the radii $r$ are drawn from a flat
distribution between either $0.49<r<0.51$ or $0.4<r<0.6$. The
masses are proportional to the radii cubed. In the former case
of almost monodisperse particles, the particles tend to
crystallize into a triangular lattice (Sec.~\ref{subsec.crystal}), 
whereas the more polydisperse particles lead to amorphous packings 
such as shown in Fig.~\ref{fig.weights}a. This allows us to study how the
packing geometry affects the force network. The results shown in 
this paper are obtained with particles that deform $0.1\%$ 
under their own weight. Simulations of harder
particles (deformation $0.01\%$) gave similar results as those
shown here \cite{foothard}. 

The various data were obtained from $1100$ realizations
containing $1180$ particles each. We study the force and
weight distributions at various heights $h$. To do so, we
divide each packing into horizontal slices of one particle
diameter thickness, and rescale all forces and weights in each
layer to the corresponding average (absolute) values. The
rescaled interparticle forces and weights will be denoted by
$\vec{f}$ and $w$ respectively, with distributions
$P(\vec{f})$ and ${\cal P}(w)$.

\subsection{Absolute values of $\vec{f}$: $P(f)$}\label{subsec.pf}
\begin{figure}[t]
\includegraphics[width=8.5cm]{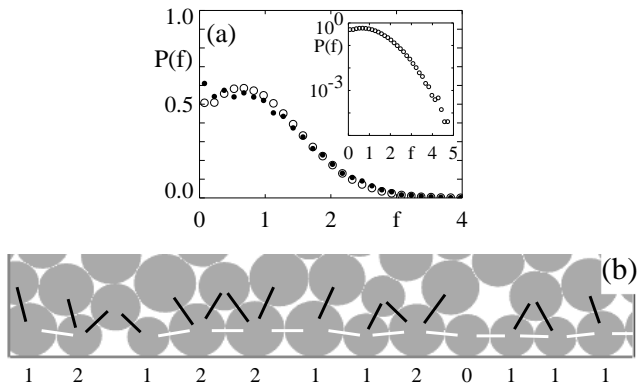}
\caption{{\em (a)} $P(f)$ for amorphous packing in the bulk (open
circles) and for the layer-to-layer forces near the bottom
(dots); the inset shows $P(f)$ for bulk forces on a log-lin scale. 
Note that the force distributions are very similar, except for 
a small difference for small $f$. 
{\em (b)} Detail of a typical packing near the bottom showing 
layer-to-layer forces (black lines) and the intralayer forces 
(white lines) near the bottom. It is clear that the layer-to-layer 
forces are dominant in determining the weights $w$ of the bottom 
particles. The numbers show the values of $n_c$, the number of 
(layer-to-layer) forces that contribute to these 
weights.} \label{fig.pfamorph}
\end{figure}

We first analyze the statistics of the absolute values
$f=|\vec{f}|$, whose probability density function $P(f)$ is
usually referred to as the distribution of (interparticle)
forces; our main finding will be that $P(f)$ in bulk and near
the boundary are very similar. In Fig.~\ref{fig.pfamorph}a we
show $P(f)$ as measured in the bulk of the {\em amorphous}
packings (particle radii between $0.4<r<0.6$). At different
heights between $10<h<30$, $P(f)$ was not observed to change; 
the open circles represent an average over these various
heights. Even very close to the bottom support, we find that
$P(f)$ remains almost unchanged: the dotted dataset has been
obtained from the forces between the bottom particles and the
particles in the layer above. We refer to these forces as
{\em layer-to-layer} forces near the bottom -- see 
Fig.~\ref{fig.pfamorph}b). So, although the bottom wall locally
alters the packing geometry, the shape of $P(f)$ is
essentially unaffected.

As can be seen from the inset of Fig.~\ref{fig.pfamorph}, 
the probability density decays slightly faster than exponentially. 
This is consistent with simulations by Makse {\em et al.} \cite{makse} 
who found that $P(f)$ crosses over to a Gaussian for large particle 
deformations; 
we have used rather `soft' particles in our simulations for which 
deformations are relatively large, i.e. up to $2\%$. 
We come back to the effect of deformation in experiments in 
Sec.~\ref{subsec.rubber}. 
For small forces, $P(f)$ approaches a finite value. 
The small peak around $f=0.7$ for bulk forces becomes a plateau
for the layer-to-layer forces near the bottom; 
it is intruiging to note that this change is reminiscent of 
what is proposed as an identification of the jamming transition 
\cite{liuletter}.

\subsection{Orientations of $\vec{f}$
and $P'(f_z)$}\label{subsec.orientation}

\begin{figure}[t]
\includegraphics[width=8.0cm]{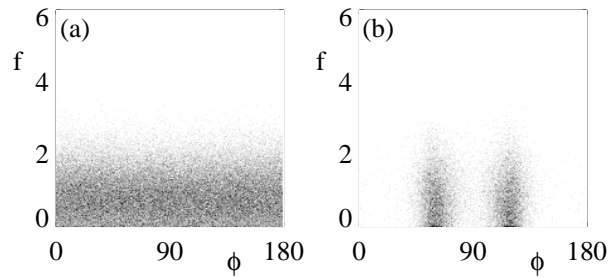}
\caption{Scatter plot of $(f_{ij},\varphi_{ij})$ for {\em (a)} 
the bulk forces, and {\em (b)} the layer-to-layer forces near the bottom 
in the amorphous packings.} \label{fig.angles}
\end{figure}

After studying the absolute values of $\vec{f}_{ij}$, let us
investigate the {\em orientations} of the interparticle forces. 
We therefore define $\varphi_{ij}$ as the angle between $\vec{f}_{ij}$ 
and the horizontal axis. 
In Fig.~\ref{fig.angles}a we show the scatterplot of 
$(f_{ij},\varphi_{ij})$ in the bulk: 
the angles are uniformly distributed and independent of the absolute 
value of $\vec{f}$. 
So, the packings are highly disordered away from the bottom. 
Near the boundary, however, this isotropy is broken
strongly. The presence of the bottom wall aligns the bottom
particles and as a consequence their 
interparticle forces become almost purely horizontal, see
Fig.~\ref{fig.pfamorph}b. It is clear that near the bottom the
interparticle forces naturally divide up into these almost
horizontal {\em intralayer} forces, and {\em layer-to-layer}
forces connecting bottom particles with those in the layer
above. The orientations of these layer-to-layer forces are 
indeed concentrated around $\pi/3$ and $2\pi/3$, as can be 
seen from Figs.~\ref{fig.pfamorph}b and~\ref{fig.angles}b.

\begin{figure}[t]
\includegraphics[width=6.0cm]{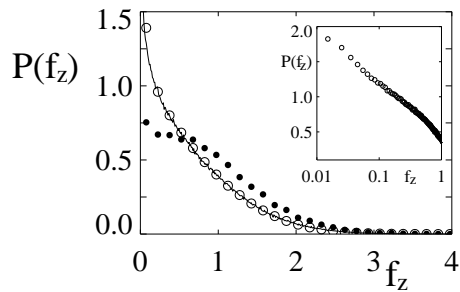}
\caption{$P'(f_z)$ in the bulk (open circles) and for the
layer-to-layer forces (dots). The solid line was obtained by
numerical integration of Eq.~(\ref{pfzdiv}). Inset shows 
$P'(f_z)$ versus log $f_z$, confirming the logarithmic 
divergence for small $f_z$.} \label{fig.pfz}
\end{figure}

Since the particle {\em weights} are derived from the
$z$-components of the forces,
$f_z=\left(\vec{f}_{ij}\right)_z$, we now investigate their 
distribution $P'(f_z)$. The bottom-induced orientational order
discussed above is reflected in the statistics of the $f_z$.
According to Fig.~\ref{fig.pfz}, there is a substantial
difference between $P'(f_z)$ in the bulk (open circles) and
$P'(f_z)$ for the layer-to-layer forces near the bottom (dots).
This difference can be understood as follows. Assuming that
the $\varphi_{ij}$ are indeed uncorrelated to the $f_{ij}$, we
can write
\begin{equation}\label{pfztrafo}
P'(f_z)=\int_0^{\pi}d\varphi \,\Phi(\varphi)\int_0^\infty df
\,P(f)\, \delta\left(f_z-f\sin(\varphi)\right),
\end{equation}
where $\Phi(\varphi)$ is the angle distribution, and $P(f)$
is the distribution of the absolute values $|\vec{f}|$ of
Fig.~\ref{fig.pfamorph}. Note that $\langle f_z \rangle < 1$.
For the layer-to-layer forces near the bottom, we have seen
from the scatter plot that the values of $\sin(\varphi)$ are
concentrated around $\frac{1}{2}\sqrt{3}\approx 0.866$. In the
approximation that the distribution of $\sin(\varphi)$ is
sharply peaked, the shape of $P'(f_z)$ equals that of $P(f)$
(up to a scale factor). This is indeed confirmed by direct
comparison of the dotted datasets of Figs.~\ref{fig.pfamorph}
and~\ref{fig.pfz}.

In the bulk, we have seen that the packing geometry is
isotropic. A consequence of this isotropy is that the
probability density function of the horizontal components
$P'(f_x)$ is identical to $P'(f_z)$ (not shown here). Again, one
can use Eq.~(\ref{pfztrafo}) to understand the shape of
$P'(f_z)$. Taking a uniform angle distribution
$\Phi(\varphi)=1/\pi$, we obtain (Appendix~\ref{app.logdiv})
\begin{equation}\label{pfzdiv}
P'(f_z)=\frac{2}{\pi}\int_{f_z}^\infty df
\,\frac{P(f)}{\sqrt{f^2-f_z^2}}.
\end{equation}
Numerical integration of this equation with $P(f)$ from
Fig.~\ref{fig.pfamorph} yields the solid line in
Fig.~\ref{fig.pfz}, which closely corresponds to the
$P'(f_z)$ as measured in the bulk (open circles). In
Appendix~\ref{app.logdiv}, we show that the integral of
Eq.~(\ref{pfzdiv}) is weakly divergent for small $f_z$:
\begin{equation}\label{pfzlog}
P'(f_z) = -\frac{2}{\pi} P(0)\, \ln(f_z) + {\cal O}(1)~.
\end{equation}
The inset of Fig.~\ref{fig.pfz} shows that our data for $P'(f_z)$ is
indeed consistent with this logarithmic divergence.

\subsection{$P(f)$: summary}\label{subsec.pfconclusion}

Let us briefly summarize the results of this section. The geometrical
constraint imposed by the bottom wall locally induces a packing
geometry which is different from the bulk geometry. Whereas this is
strongly reflected in the orientations of the $\vec{f}_{ij}$, the
distribution of the absolute values $P(f)$ is very robust. The
probabilities for the components of the $\vec{f}_{ij}$ can be obtained
with great precision, including the logarithmic divergence, by the
transformation of Eq.~(\ref{pfztrafo}). 

\section{Packing geometry and weight distributions 
${\cal P}(w)$}\label{sec.pw}

In this section, we demonstrate that the local packing geometry has a
dramatic effect on the weight distribution of ${\cal P}(w)$. As
stated in the introduction, experiments can only measure the
particle-wall forces at the boundary of a granular packing, and not
the interparticle (bulk) forces that were discussed in the previous
section. Since these particle-wall forces are essentially equal to the
weights of the bottom particles, it is important to understand the
relation between the weight distribution ${\cal P}(w)$ and the
distribution of interparticle forces $P(\vec{f})$. In the first part
of this section we develop a simple geometrical framework to
understand this relation, based on phase space considerations. We
then show that this explains, to a large extent, the weight
distributions ${\cal P}(w)$ as measured in our simulations of Hertzian
spheres. In particular, we observe substantial differences between
weight distributions for different packing geometries.

\subsection{Geometrical framework: decomposition of ${\cal P}(w)$ 
according to number of contacts $n_c$ from above}\label{subsec.framework}
 
If we interpret Eq.~(\ref{defweight}) as a transformation of stochastic
variables, it is possible to relate the corresponding probability
density functions as
\begin{eqnarray}\label{trafo}
{\cal P}_{n_c}(W)&=&\int_0^\infty d(\vec{F}_{1})_z \cdots
\int_0^\infty d(\vec{F}_{n_c})_z \, \nonumber \\
&\times&
P\left((\vec{F}_{1})_z,\cdots,(\vec{F}_{n_c})_z\right)
\, \delta\left(W-\sum_{i=1}^{n_c} (\vec{F}_{i})_z \right)~. \nonumber \\
&&
\end{eqnarray}
Here, we have neglected the term $mg$, since 
$mg /\langle W \rangle \ll 1$ far below the top surface of the packing. 
The number of forces over which we integrate differs from grain 
to grain, and it turns out to be crucial to label the weight 
distribution in Eq.~(\ref{trafo}), ${\cal P}_{n_c}(W)$, 
according to this number $n_c$. This can be seen as follows. 
The $\delta$-function constrains the integral on a $(n_c-1)$
dimensional hyperplane of the total phase space, and the
``area'' of this hyperplane scales as $W^{n_c-1}$. We thus
anticipate the following scaling behavior for small weights:
\begin{equation}\label{pncscaling}
{\cal P}_{n_c}(W)\propto W^{n_c-1} \quad{\rm for} \quad
w\rightarrow0~,
\end{equation}
provided that the joint probability density approaches a finite 
value when all $(\vec{F}_{i})_z \rightarrow 0$. 
Such scaling is also implicit in the $q$-model \cite{qm}, although 
there $n_c\geq 2$ so that ${\cal P}(0)=0$. The particles
that do not feel a force from above, $n_c=0$, give a
$\delta$-like contribution at $W=mg$; for deep layers this
occurs for $mg/\langle W\rangle\ll1$. 
In a disordered packing, the number of
particles that exert a force from above can vary from grain to
grain. The total weight distribution ${\cal P}(W)$, therefore,
is a superposition of the ${\cal P}_{n_c}(W)$:
\begin{equation}\label{pwdecomp}
{\cal P}(W)=\sum_{n_c} \rho_{n_c}\, {\cal P}_{n_c}(W)~,
\end{equation}
where $\rho_{n_c}$ is the fraction of particles with $n_c$
contacts from above. This means that the small weight behavior
of ${\cal P}(W)$ depends very much on the fractions
$\rho_{n_c}$ and thus on the local packing geometry, via
Eqs.~(\ref{pncscaling})~and~(\ref{pwdecomp}).

The steepness of the tail of the total weight distribution
depends strongly on the $\rho_{n_c}$ as well. To explain this,
let us assume that all vertical forces $F_z$ contributing to
the weight are uncorrelated. We consider $P'(f_z)\propto
e^{-\alpha f_z}$, i.e. $P'(F_z)\propto e^{-\alpha F_z/\langle F_z \rangle}$ 
for large forces. 
It follows from Eq.~(\ref{trafo}) that the
weight distribution takes over this same exponent 
$\alpha/\langle F_z \rangle$, so that ${\cal
P}_{n_c}(W)\propto e^{-\alpha W/\langle F_z \rangle}$. However, the 
${\cal P}_{n_c}(W)$'s are not properly normalized: $\langle W
\rangle_{n_c}=\langle F_z \rangle \,n_c$, since each of the
$F_z$ gives an average contribution $\langle F_z \rangle$.
This yields a total average weight $\langle W \rangle=\langle F_z\rangle
\sum_{n_c} \rho_{n_c} n_c=\langle F_z\rangle\,\langle n_c\rangle$. 
In order to compare with experimental and theoretical results 
we have to rescale the weights so that $\langle w\rangle =1$, 
yielding the following large weight behavior:

\begin{equation}\label{pwtails}
{\cal P}(w)\propto e^{-\gamma w} \quad{\rm with}\quad
\gamma=\alpha \, \langle n_c \rangle~.
\end{equation}
This simple calculation shows that, for a given value of
$\alpha$, the steepness of the tail of the experimentally measured 
weight distribution is very sensitive to the local packing geometry.
This is a direct consequence of keeping $\langle w \rangle$
fixed to unity: a decrease of probability for small weights
must lead to a steeper tail for large weights in order to
leave the average weight unaltered. Note that this general
argument is not restricted to uncorrelated $F_z$ or
exponential tails. A generalization to other than exponential
tails is given in Appendix~\ref{app.tails}.

So, we have advanced a simple picture, in which the shape of
${\cal P}(w)$ {\em depends strongly on the local packing
geometry} via the fractions $\rho_{n_c}$. The small force
behavior follows from Eqs.~(\ref{pncscaling})
and~(\ref{pwdecomp}), whereas Eq.~(\ref{pwtails}) relates to a
good approximation the exponential tails of $P'(f_z)$ and
${\cal P}(w)$. The object one ultimately wishes to
characterize is of course the force distribution $P(f)$. Since
close to the boundary $P(f)$ and $P'(f_z)$ are identical up to
a scaling factor $\langle f_z \rangle$ (Sec.~\ref{subsec.orientation}), 
the above equations allow to trace the features of the force 
distribution from experimental measurements. Along this line, 
we analyze recent experimental data in Sec.~\ref{subsec.rubber}.

\subsection{${\cal P}(w)$ in Hertzian sphere packings}
\begin{figure}[t]
\includegraphics[width=8.5cm]{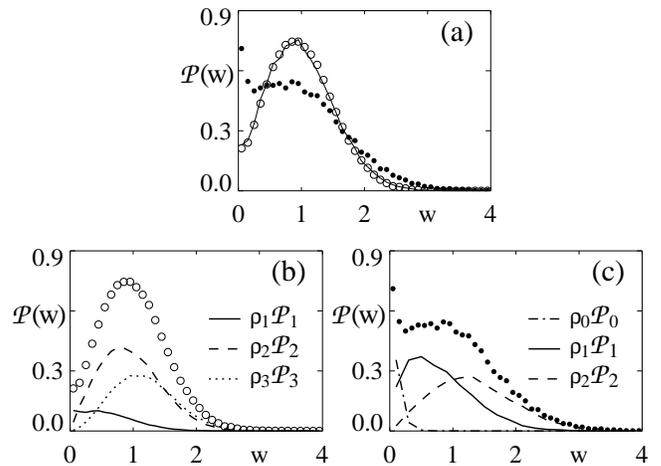}
\caption{{\em (a)} ${\cal P}(w)$ in the bulk (open circles) and 
at the bottom (dots) in amorphous packings. 
At $2<h<3$, ${\cal P}(w)$ is already bulk-like (solid line). 
{\em (b,c)} Decomposition of ${\cal P}(w)$ according to 
Eq.~(\ref{pwdecomp}) {\em (b}) in the bulk (open circles) and 
{\em (c)} at the bottom (dots). The measured bulk values for the
 fractions $\{\rho_0,\rho_1,\rho_2,\rho_3\}$ in Eq.~(\ref{pwdecomp}) 
are $\{0.01,0.11,0.52,0.36\}$, and the bottom values are
 $\{0.08,0.46,0.44,0.02\}$; as explained in \cite{footnc}, we excluded 
the intralayer (almost horizontal) forces at the bottom when 
determining $n_c$.} \label{fig.pwamorph}
\end{figure}

We now discuss the weight distributions observed in the
Hertzian sphere packings, and interpret the results within the
framework developed above. Figure~\ref{fig.pwamorph}a shows that
in the amorphous packing ${\cal P}(w)$ in the bulk (open
circles) is significantly different from ${\cal P}(w)$ of the
bottom particles (dots). The probability for small weights is
much larger at the bottom, and the decay for large weights is
not as steep as for the bulk particles. Furthermore, the
transition from bottom to bulk behavior is remarkably sharp:
in the slice $2<h<3$ (full curve), the weight distribution is
already bulk-like.

\begin{figure}[t]
\includegraphics[width=8.5cm]{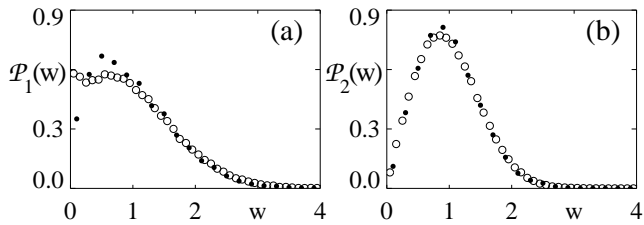}
\caption{Direct comparison of {\em (a)} ${\cal P}_1(w)$ and {\em (b)}
${\cal P}_2(w)$ for bulk (open circles) and bottom particles
(dots). All distributions are scaled such that $\langle w
\rangle=1$.} \label{fig.pncw}
\end{figure}

Using the concepts developed in the preceding paragraphs, we now 
show how this change in ${\cal P}(w)$ can be explained by a 
change in the local packing geometry. 
Consider the typical bottom
configuration of Fig.~\ref{fig.pfamorph}b. The {\em intralayer
forces} (white lines) are almost purely horizontal and hence do 
not contribute to the weights. This reduces the effective values
of $n_c$, leading to the following fractions for the bottom
particles: 
$\{\rho_0,\rho_1,\rho_2,\rho_3\}=\{0.08,0.46,0.44,0.02\}$, 
where we did not count the intralayer forces for determining 
the values of $n_c$ \cite{footnc}. 
In the bulk, these fractions are different, namely
$\{\rho_0,\rho_1,\rho_2,\rho_3\}=\{0.01,0.11,0.52,0.36\}$. 
According to Eq.~(\ref{pwdecomp}), these differences between 
the $\rho_{n_c}$ in the bulk and at the bottom should lead to a 
substantially different ${\cal P}(w)$. 
Figs.~\ref{fig.pwamorph}b,c explicitly shows the decomposition
into the ${\cal P}_{n_c}(w)$. Indeed, one observes the scaling
behavior for small $w$ proposed in Eq.~(\ref{pncscaling}). Moreover, 
the various ${\cal P}_{n_c}(w)$ are essentially the same at 
the bottom and in the bulk: a direct comparison is given in
Fig.~\ref{fig.pncw}, where we rescaled the average values to
unity. There is only a small difference in the ${\cal P}_1(w)$
due to the fact that bottom particles with $n_c=1$ are
typically smaller than average (Fig.~\ref{fig.pncw}a). For
these particles, the intralayer forces will add a small
contribution to the weights, enhancing ${\cal P}_1(w)$ for
small $w$ at the expense of ${\cal P}_1(0)$. The same argument
holds for ${\cal P}_0(w)$, whose $\delta$-like shape appears a
bit broadened in Fig.~\ref{fig.pwamorph}c.
However, it is clear that the differences between ${\cal P}(w)$ 
in the bulk and at the bottom are mainly due to a change in 
contact geometry.

Finally, let us remark that the good agreement between $P_{bulk}(f)$ 
and ${\cal P}_{boundary}(w)$ for $w>0.3$ is fortuitous and due to 
the relatively large fraction of bottom particles with $n_c=1$. 
We will argue below that this is also the case in many 
(but not all) carbon paper experiments.

\subsection{Summarizing the simple picture}

Our simple framework as developed in the sections above can be
summarized as follows: The geometry of the contact network has a
strong effect on ${\cal P}(w)$, while $P(f)$ is very robust. The weight
distribution for particles with a given $n_c$, ${\cal P}_{n_c}({w})$,
is robust and behaves as $w^{n_c-1}$ for small $w$. ${\cal P}(w)$ can
be decomposed as ${\cal P}({w})=\sum_{n_c} \rho_{n_c} {\cal
P}_{n_c}({w})$, where $\rho_{n_c}$ are the fractions of particles that
have $n_c=0,1,2,\dots$ ``up'' contacts. Differences of $\rho_{n_c}$
between boundary particles and bulk particles explain the different
${\cal P}(w)$'s for these cases. When $\rho_0$ and $\rho_1$ are large,
the total weight distributions ${\cal P}(w)$ exhibits a plateau at
small weights and a slow decay at large weights; when $\rho_2$ and
$\rho_3$ become large, ${\cal P}(w)$ becomes sharply peaked. In this
way, ${\cal P}(w)$'s small weight behavior as well as its exponential
decay rate for large weights reflect the packing geometry.

\section{Manipulating the geometry: experimental relevance}\label{sec.exp}

So far we have focused on the role of the bottom boundary for 
disordered packings of frictionless particles. 
In this section we provide explicit examples of other types of packing 
geometries and their effect on ${\cal P}(w)$. We first discuss our 
simulations of weakly polydisperse particles, which give rise to rather 
crystalline packings -- see Fig.~\ref{fig.crystal}a. We then apply the 
geometrical framework derived in the previous section to 
experimental (carbon paper) data by Erikson {\em et al.} \cite{rubber} 
of highly deformed packings of soft rubber particles. Their results have 
a natural interpretation within our framework and form a nice illustration 
of how the number of contact affects the weight distribution. 
Both the simulations of crystalline packings and the experiments 
on deformed packings are examples where the experimentally accessible 
${\cal P}_{boundary}(w)$ is significantly different from $P(f)$ in 
the bulk; we discuss why in many other carbon paper experiments 
${\cal P}_{boundary}(w)$ is probably very similar to the real $P(f)$.

\subsection{Crystalline versus disordered frictionless packings}
\label{subsec.crystal}

\begin{figure}[t]
\includegraphics[width=8.5cm]{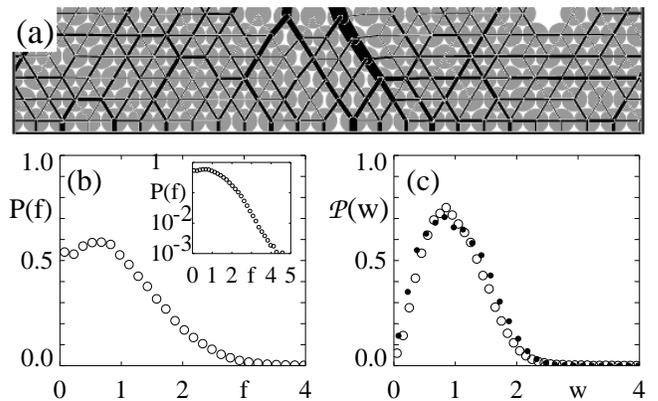}
\caption{{\em (a)} Weakly polydisperse particles (radii between $0.49<r<0.51$) 
spontaneously crystallize into a hexagonal packing. 
{\em (b)} The corresponding $P(f)$ is indistinguishable from 
the force distributions in amorphous packings. 
{\em (c)} The weight distributions ${\cal P}(w)$ in the bulk (open circles) 
and at the bottom (dots) are dominated by particles with $n_c=2$.}
\label{fig.crystal}
\end{figure}

We now present the results of the more or less crystalline packings, 
obtained from simulations with particle radii between $0.49<r<0.51$. 
Firstly, the force distribution $P(f)$ shown in Fig.~\ref{fig.crystal}b is 
indistinguishable from the force distributions in the amorphous packings 
(compare with Fig.~\ref{fig.pfamorph}a). 
So despite the order in particle positions, there are still large 
fluctuations in the force network. There is of course some disorder 
in the ``contact network'' since not all particles are in contact 
with their six neighbors (Fig.~\ref{fig.crystal}a). 
It is nevertheless surprising that for this very different contact 
geometry, the force fluctuations are characterized by the same 
probability distribution as was observed for highly disordered packings. 
This strongly suggests that $P(f)$ is a very 
robust quantity and independent of the packing geometry. 

The weight distribution ${\cal P}(w)$, on the other hand, is very 
sensitive for the geometry. In a perfect triangular packing all 
particles would have $n_c=2$; in our simulations we find that 
$\rho_2=0.9$ and $\rho_1=0.1$ due to lattice imperfections. 
From our geometrical framework we expect that the shape of the weight 
distribution is dominated by ${\cal P}_2(w)$. Fig.~\ref{fig.crystal}c 
shows that this is indeed the case -- 
e.g. compare with Fig.~\ref{fig.pncw}b. 

In an earlier paper \cite{heggn}, we reported how one can break the 
regular packing geometry by using curved boundaries. This led to a 
dramatic change in ${\cal P}(w)$ that again could be understood 
from a change in the $\rho_{n_c}$. 

\subsection{Experiments on strongly deformed particles}\label{subsec.rubber}

We now demonstrate how the strategy to decompose the weight 
distributions according to $n_c$ can be applied to experiments 
measuring ${\cal P}(w)$ at the boundary of a granular material. 
This is best illustrated by recent carbon paper experiments by 
the Chicago group on soft rubber beads, in particular Fig.~3 of 
Ref.~\cite{rubber}, 
in which the effect of particle deformations was investigated. 
The raw data of these experiments were kindly made available by 
the authors, allowing us to perform the analysis presented below.

The experimental results of Fig.~3 of Ref.~\cite{rubber} display 
three trends as the compression is increased:
\begin{itemize}
\item The $\delta$-like peak at $w=0$ decreases,
\item $\lim_{w \downarrow 0} {\cal P}(w)$ decreases,
\item The exponential tail becomes steeper.
\end{itemize}
These behaviors emerge naturally when considering the role of the 
fractions $\rho_{n_c}$. 
The first trend arises from a decrease in $\rho_0$, since only 
particles with $n_c=0$ give a $\delta$-like contribution to 
${\cal P}(w)$. The second trend comes from a decrease in $\rho_1$: 
from Eqs.~(\ref{pncscaling}) and~(\ref{pwdecomp}) it is clear that 
$\lim_{w \downarrow 0} {\cal P}(w)=\rho_1 {\cal P}_1(w)$. 
The changes in ${\cal P}(w)$ can thus be understood from an increasing 
number of contacts, which is what one would expect for a 
compressed system \cite{makse}. 
The fractions $\rho_2$ and $\rho_3$ will increase at the expense of 
$\rho_0$ and $\rho_1$. 
Also the third trend, 
the steepening of the exponential tail, is directly related 
to the increase in $\langle n_c \rangle$ via Eq.~(\ref{pwtails}). 
However, Eqs.~(\ref{pncscaling})-(\ref{pwtails}) allow to further 
quantify this change in contact geometry from the experimental data. 
The value of $\rho_1{\cal P}_1(0)$ can be read off from the plots, 
after subtracting the $\delta$-like data points, 
since $\rho_1 {\cal P}_1(0)=\lim_{w \downarrow 0} {\cal P}(w)$. 
The value of $\rho_0$ is obtained by the height of the $\delta$-peak 
times the bin-width. Using the raw experimental data, 
we obtained the figures given in the first colomn of Table~\ref{tabrubber}, 
where we took ${\cal P}_1(0)=0.5$ \cite{foottails}. 
Unfortunately, the values of $\rho_2$ and $\rho_3$ can not be 
determined directly from the data. 

An intriguing issue is that numerical simulations by 
Makse {\em et al.} \cite{makse} indicate that $P(f)$ crosses over to 
a Gaussian for large particle deformations. 
This contradicts the experimental data for which one observes an 
exponential tail even though particle deformations are up to no 
less than $45\%$ \cite{rubber}. 
Moreover, we speculate below that the steepening of the tails is 
only due to changes in the $\rho_{n_c}$, and that the bulk force 
distributions $P(f)$ actually remain unaffected by the particle 
deformations. 
The way to test this scenario is to examine whether the 
exponential decay constant of $P(f)\propto e^{-\hat{\alpha}f}$ 
remains fixed, even though the steepness of 
${\cal P}(w)\propto e^{-\gamma w}$ increases. 
We use Eq.~(\ref{pwtails}) to determine the value of 
$\alpha =\gamma/\langle n_c \rangle$, 
where $\alpha$ and $\gamma$ are the decay rates of $P'(f_z)$ and 
the experimental ${\cal P}(w)$ respectively. 
Since we found in Sec.~\ref{subsec.orientation} that 
$P(F)$ and $P'(F_z)$ near the bottom are almost identical up to 
a scaling factor $\langle F_z \rangle/\langle F\rangle$, 
the actual decay rate of $P(f)\propto e^{-\hat{\alpha}f}$ is 
exactly the same as that of the (renormalized) $P'(f_z)$, so that 
$\hat{\alpha}=\alpha$. Hence, we can approximate the exponential 
decay constant of the force distribution as
\begin{equation}
\hat{\alpha}=\frac{\gamma}{\langle n_c \rangle}~.
\end{equation}
To estimate the values of $\langle n_c \rangle$, we worked out 
two scenarios: we take either $\rho_2=\rho_3$ or $\rho_3=0$. 
Together with the values of $\rho_0$, $\rho_1$ and $\gamma$, 
taken from the experimental data, 
this yields the values of $\hat{\alpha}$ listed in the second 
and third column of Table~\ref{tabrubber}. Surprisingly, 
the root mean square deviation in $\hat{\alpha}$ is only $18 \%$, 
which is rather small considering our rather crude estimates of the 
$\rho_{n_c}$ and the fact that Eq.~(\ref{pwtails}) is only approximate. 

Let us briefly recapitulate the discussion above. 
First, we have interpreted the changes in experimental particle-wall 
force distributions of strongly compressed packings \cite{rubber} 
as a change in the packing geometry. To be more precise, the overall 
trends can be understood from the expected increase of the number of 
contacts due to compression. 
We demonstrated how one can determine the fractions $\rho_0$ and $\rho_1$ 
from the experimental data. 
Direct measurements of these fractions would be very welcome as a test 
of our framework, as well as to extract further information of the force 
distribution $P(f)$. 
Furthermore, our crude estimates in Table~\ref{tabrubber} give reason to 
believe that the force distribution $P(f)$ is actually not much affected 
by the compression. Finally, it seems that for most experimental results, 
where particle deformations are relatively small, $\rho_0$ and $\rho_1$ 
are substantial at the boundary, so that ${\cal P}_{boundary}(w)$ is 
similar to $P_{bulk}(f)$ (apart from a $\delta$-peak at $w=0$). 
The same argument probably holds for recent simulations by 
Silbert {\em et al.} \cite{silbert}.

\begin{table}[ht]
\begin{center}
\begin{tabular}{c|ccc|cc|cc} 
\hline
 && &&$\rho_2=\rho_3$ &  & $\rho_3=0 $ &  \\*[2mm]
 & & & & & & &\\*[-3mm]
deform. & $\gamma$ & $\rho_0$ & $\rho_1$ &  
$\langle n_c \rangle$ & $\hat{\alpha}=\frac{\gamma}{\langle n_c \rangle}$ &
$\langle n_c \rangle$ & $\hat{\alpha}=\frac{\gamma}{\langle n_c \rangle}$ 
\\*[3mm]
\hline
$25\%$ & 2.4 & 0.23 & 0.58 & 1.05 & 2.29 & 0.96 & 2.51 \\*[2mm]
  & & & & &&&\\*[-3mm]
$30\%$ & 2.6 & 0.21 & 0.26 & 1.60 & 1.63 & 1.33 & 1.96 \\*[2mm]
  & & & & &&&\\*[-3mm]
$37\%$ & 2.8 & 0.14 & 0.18 & 1.88 & 1.49 & 1.54 & 1.81 \\*[2mm]
  & & & &&& &\\*[-3mm]
$45\%$ & 3.8 & 0.00 & 0.05 & 2.42 & 1.57 & 1.95 & 1.95 \\*[2mm]
\hline
\end{tabular} 
\centering
\caption{The calculated values for the exponents $\hat{\alpha}$, 
after estimating the fractions $\rho_{n_c}$ 
from the experimental data of Figs.~3a-d of Ref.~\cite{rubber}. 
The percentage in the first column represent the degree of 
particle deformation. 
The values of $\gamma$ are taken from Table I of Ref.~\cite{rubber}.}
\label{tabrubber}
\end{center}
\end{table}

\section{Beyond the simple picture}\label{beyond}

In the picture that we have constructed above we characterize 
the packing geometry by the fractions $\rho_{n_c}$, and we found 
that the ${\cal P}_{n_c}(w)$ are very robust. 
This is of course a vast simplification, since we characterize 
the local environment of a particle by only one number, namely $n_c$. 
In this section we address the question why this crude approach 
works so remarkably well. 
For bottom particles the situation is particularly simple and 
insightful, since the geometry of the contacts is more or less 
fixed. There is one contact with the bottom, one or two almost 
horizontal intralayer contacts and $n_c$ forces from above -- 
Fig.~\ref{fig.pfamorph}b. As we have shown in Fig.~\ref{fig.angles}b, 
the angles of these forces display little scatter, so the local 
texture is more or less fixed once $n_c$ is given. For bottom particles 
one can thus understand that $n_c$ indeed provides a good description 
of the local packing geometry, which justifies the decompostion 
according to $n_c$. Although for particles in the bulk the situation 
is more complicated, there are similar arguments why ${\cal P}_{n_c}(w)$ 
is indeed a robust quantity, i.e. insensitive for packing geometry. 
These will be discussed in Sec.~\ref{subsec.pncrobust}. 
We then address the up-down symmetry of the system. Our framework 
only involves the number of contacts from above, $n_c$, and not the 
number of contacts from below, $n_b$. For bottom 
particles $n_c$ is the obvious parameter, but in the bulk of an 
amorphous packing, 
where the angle distribution is isotropic, there is no reason why 
$n_c$ should be more important than $n_b$. In Sec.~\ref{subsec.pncnb} 
we therefore investigate weight distributions for particles with a 
given combination $\{n_c,n_b\}$, which we denote by 
${\cal P}_{n_cn_b}(w)$. 
Special attention will be paid to particles that have $n_c \neq n_b$ 
in Sec.~\ref{subsec.ncneqnb}.

\subsection{Why is ${\cal P}_{n_c}(w)$ for bulk particles robust?}
\label{subsec.pncrobust}
It is not a priori clear why ${\cal P}_{n_c}(w)$ is rather 
insensitive for the packing geometry, since the definition of 
${\cal P}_{n_c}(w)$ in Eq.~(\ref{trafo}) 
involves the joint distribution of the $(\vec{f}_{i})_z$ that push 
on a particle from above, i.e. 
$P\left((\vec{f}_{1})_z,\cdots,(\vec{f}_{n_c})_z\right)$. 
This joint distribution has an explicit geometry dependence since the 
projections in the $z$-direction involve the distribution of 
contact angles $\varphi_i$. 
Even if we assume that the force {\em magnitude} is 
uncorrelated to its {\em orientation}, i.e. 
\begin{equation}
P\left(\vec{f}_1, \cdots, \vec{f}_{n_c}\right) =
P\left(f_1, \cdots, f_{n_c}\right) 
\Phi(\varphi_1,\cdots,\varphi_{n_c}) \,~,
\end{equation}
we obtain the distribution of the vertical components 
$P\left((\vec{f}_{1})_z,\cdots,(\vec{f}_{n_c})_z\right)$ 
by integration over the joint angle distribution 
$\Phi(\varphi_1,\cdots,\varphi_{n_c})$. 
Therefore, the ${\cal P}_{n_c}(w)$ have an explicit geometry 
dependence. 

\begin{figure}[t]
\includegraphics[width=8.5cm]{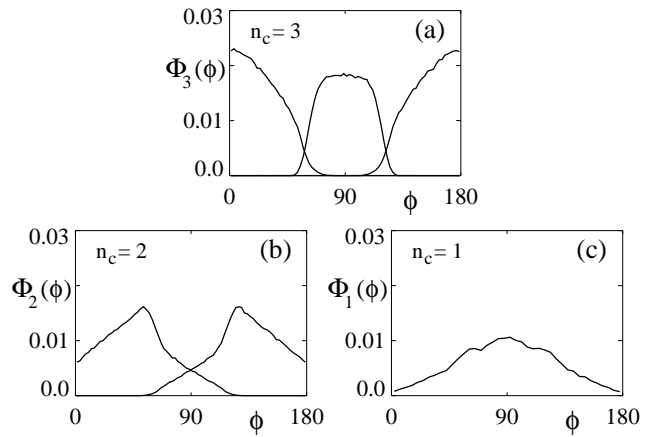}
\caption{{\em (a)} For particles with $n_c=3$, we plot the probability 
densities for the angles $\Phi_3(\varphi_1)$, $\Phi_3(\varphi_2)$ and 
$\Phi_3(\varphi_3)$, where the three angles have been sorted such that 
$\varphi_1 < \varphi_2 < \varphi_3$; 
{\em (b)} The probability densities $\Phi_2(\varphi_1)$ and $\Phi_2(\varphi_2)$ 
for particles with $n_c=2$; 
{\em (c)} The probability density $\Phi_1(\varphi_1)$ for particles 
with $n_c=1$.}
\label{fig.phinc}
\end{figure}

We already saw that this angle distribution is more or 
less fixed for bottom particles. 
For the polydispersities used in 
this study, the bulk angles have also limited room for 
fluctuations once $n_c$ has been specified. For example if $n_c=3$, 
one typically finds one angle close to $\pi/2$ and two relatively 
small angles, see Fig.~\ref{fig.phinc}a; this is because the three particles 
should all touch the upper half of the bead supporting them. 
Particles with $n_c=2$ also have such an ``excluded 
volume''-like constraint (Fig.~\ref{fig.phinc}b), albeit less strong than for
$n_c=3$. Particles with $n_c=1$ have an enhanced probability for
angles around $\pi/2$, because such contacts make the presence of a
second contact from above less probable (Fig.~\ref{fig.phinc}c). 
So, the shape of ${\cal P}_{n_c}(w)$ is limited by the geometric constraints 
on the angle distributions $\Phi(\varphi_1,\cdots,\varphi_{n_c})$, 
which are rather peaked. 
This justifies the picture that the geometry dependence of ${\cal P}(w)$ 
is mainly due to the $\rho_{n_c}$, and that the ${\cal P}_{n_c}(w)$ 
can be considered invariant. 

Note that the abovementioned constraints on the angle distributions 
imply that the averages $\langle w \rangle_{n_c}$ are not simply 
proportional to $n_c$. Comparing for example $n_c=1$ and $n_c=3$, 
we see that the two ``extra'' forces for $n_c=3$ 
have a relatively small vertical component; 
the average weight will thus grow less than linearly with 
$n_c$. We should therefore correct Eq.~(\ref{pwtails}) for the 
steepness of the tails by replacing $\langle n_c \rangle$ with 
$\sum_{n_c} \rho_{n_c} \langle w \rangle_{n_c}$. 
Making a correction of this type would further refine our 
analysis of the experiment with rubber beads discussed in 
Sec.~\ref{subsec.rubber}.

\subsection{Gravity and up-down symmetry}\label{subsec.pncnb}
In our analysis of ${\cal P}(w)$ we have explicitly broken the
up-down symmetry, since it only involved the number of
contacts from above. At the bottom, this is an obvious choice. 
Away from the boundary, however, the amorphous packings have 
an isotropic angle distribution even though the packings 
were created under gravity. Moreover,
we have neglected the term $mg$ in Eq.~(\ref{defweight}),
which makes the sum of forces from below equal to the sum of
forces from above. So in principle one could also decompose 
${\cal P}(w)$ according to the number of contacts from below $n_b$. 
We therefore investigate ${\cal P}_{n_cn_b}(w)$; this can be regarded 
as a ``component'' of ${\cal P}_{n_c}(w)$, since 
$\rho_{n_c} {\cal P}_{n_c}(w)=\sum_{n_b} \rho_{n_cn_b} {\cal P}_{n_cn_b}(w)$.

\begin{figure}[t]
\includegraphics[width=8.5cm]{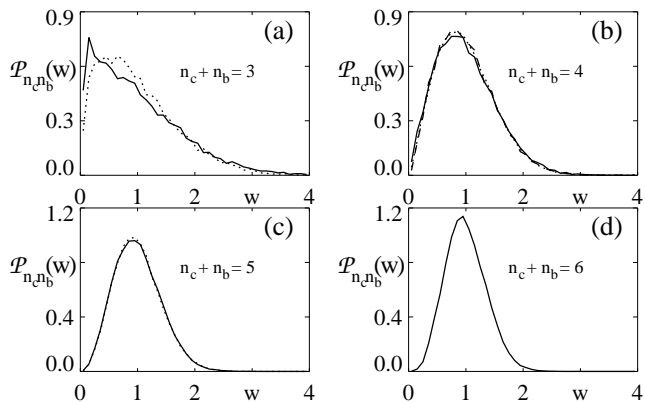}
\caption{{\em (a)} ${\cal P}_{12}(w)$ (solid line) 
and ${\cal P}_{21}(w)$ (dotted line); 
{\em (b)} ${\cal P}_{13}(w)$, ${\cal P}_{22}(w)$ and  ${\cal P}_{31}(w)$; 
{\em (c)} ${\cal P}_{23}(w)$ and ${\cal P}_{32}(w)$; 
{\em (d)} ${\cal P}_{33}(w)$.}
\label{fig.abovebelow1}
\end{figure}

Fig.~\ref{fig.abovebelow1}b shows that ${\cal P}_{13}(w)$, 
${\cal P}_{22}(w)$ and ${\cal P}_{31}(w)$ are almost identical. 
The same holds for ${\cal P}_{23}(w)$ and ${\cal P}_{32}(w)$ 
(Fig.~\ref{fig.abovebelow1}c), so the total coordination number 
$n_c+n_b$ appears to be a more fundamental quantity than just 
$n_c$ or $n_b$. 
Fig.~\ref{fig.abovebelow1}d furthermore shows that the quadratic
scaling of ${\cal P}_{33}(w)$ is somewhat more pronounced than
for ${\cal P}_{23}(w)$ and ${\cal P}_{32}(w)$; it seems that
the presence of $2$ contacts from above or below inhibits the 
pure quadratic scaling. 

The presence of gravity is noticed,
however, for ${\cal P}_{12}(w)$ and ${\cal P}_{21}(w)$ which
do show some differences (Fig.~\ref{fig.abovebelow1}a). These
particles have only 3 contacts and were less restricted during
the formation of the static force network by the ``cage''
surrounding them. This allowed gravity to influence their
final movements more than for particles with $n_c+n_b>3$.
Obviously, this effect is even stronger for particles with
only 2 contacts, which typically have $\{n_c,n_b\}=\{0,2\}$.

To further investigate the up-down symmetry, we list the
fractions $\rho_{n_cn_b}$ of particles with a certain $n_c$
and $n_b$ in Table~\ref{table}. For all particles with 3 or
more contacts these fractions are almost perfectly symmetric.
From this we conclude that in the amorphous packings, the
up-down asymmetry due to gravity is only noticed by particles
that have 2 or 3 contacts. 

\begin{table}[ht]
\begin{center}
\begin{tabular}{c|ccccc}
\hline & & & & &\\*[-3mm] $n_c$ $\setminus$ $n_b$ &
$\quad0\quad$ & $\quad1\quad$ & $\quad2\quad$ & $\quad3\quad$
& $\quad4\quad$ \\*[2mm] \hline & & & & &\\*[-3mm] $0$ & $0$
& $0$ & $0.6$ & $0$ & $0$ \\*[2mm] $1$ & $0$ & $0.3$ & $5.6$ &
$4.7$ & $0.2$ \\*[2mm] $2$ & $0$ & $4.7$ & $26.1$ & $20.5$ &
$0.7$ \\*[2mm] $3$ & $0$ & $5.1$ & $21.6$ & $8.9$ & $0$
\\*[2mm] $4$ & $0$ & $0.3$ & $0.7$ & $0$ & $0$ \\*[2mm] \hline
\end{tabular}
\caption{Fractions $\rho_{n_cn_b}$ expressed in percentages; the
numbers are almost up-down symmetric, except for rattlers (particles
with 2 contacts). From these fractions one finds the average 
coordination number $\langle n_c+n_b\rangle =4.51$.} \label{table}
\end{center}
\end{table}

\subsection{Particles with $n_c\neq n_b$}\label{subsec.ncneqnb}

We have seen that for particles with $\{ n_c,n_b\}=\{3,1\}$ or vice 
versa, the small weight behavior is $\sim w^1$, which is 
different from the scaling predicted by Eq.~(\ref{pncscaling}).  
This breakdown of our simple picture can be understood as follows. 
A particle that has 4 contacts can either have 
$\{ n_c,n_b\}=\{3,1\}$, $\{ n_c,n_b\}=\{2,2\}$ 
or $\{ n_c,n_b\}=\{1,3\}$ depending on the precise orientations of 
the forces with respect to gravity. However, if we were to define the weights 
by projecting the $\vec{F}_{ij}$ at a small angle with respect to gravity, 
a particle with 4 contacts can easily change from 
$\{ n_c,n_b\}=\{3,1\}$ to $\{2,2\}$ or even to $\{1,3\}$. 
However, we have seen that there is no ``preferred'' projection 
direction, since gravity has only very little effect on our packings. 
Hence, it is not surprising that the ${\cal P}_{n_b n_c}(w)$ depend 
on $n_c+n_b$ and not on $n_c$ or $n_b$ individually. 

But what determines the precise scaling for small weights? 
Consider a particle $i$ with $n_c=3$ and $n_b=1$. The three forces 
pushing it from above, $\vec{F}_{i1}$, $\vec{F}_{i2}$ and $\vec{F}_{i3}$, 
are not independent: force equilibrium in the direction perpendicular 
to $\vec{F}_{i4}$ (the force pushing from below) requires 
$\left(\vec{F}_{i1}+\vec{F}_{i2}+\vec{F}_{i3}\right)\cdot \vec{n}=0$, 
where $\vec{F}_{i4}\cdot \vec{n}=0$. This reduces the 
number of independent forces from above to only 2, since the third 
is determined by mechanical equilibrium. As a consequence, 
the scaling behavior for small $w$ will be ${\cal P}_{31}(w)\propto w$. 

For particles with $n_c=3$ and $n_b=2$, the 5 forces are also coupled 
through mechanical equilibrium. In this case, however, one can not 
distil a relation between the forces from above only, such as we 
did for particles with $\{n_c,n_b\}=\{3,1\}$. So one still expects 
that ${\cal P}_{32}(w)\propto w^2$, as is observed in Fig.~\ref{fig.abovebelow1}c. 
Nevertheless, this illustrates that two dimensional mechanical 
equilibrium {\em does} introduce correlations between all forces 
pushing from above. 
This limits the validity of our arguments used in Sec.~\ref{sec.pw}, 
for bulk particles. At the bottom our analysis is still valid: 
horizontal equilibrium can be accomplished by the forces between 
neighboring bottom particles (see Fig.~\ref{fig.pfamorph}b), 
so the forces from above can really be considered as independent.

\subsection{Summary}
In this section we have addressed the limitations of our 
simple geometrical framework. We have shown that the observation 
that ${\cal P}_{n_c}(w)$ is insensitive to packing geometry 
originates from excluded volume-like correlations between the 
angles at which forces press upon a bead (Fig.~\ref{fig.phinc}). 
This is the subtle underlying reason why our simple picture, where 
we characterize the local packing geometry by only one number $n_c$, 
is good enough to interpret experimental and numerical data. 
We have furthermore studied the effect of gravity by decomposing 
the weight distribution according to the number of particles 
from below ($n_b$) as well. We found that gravity breaks the up-down 
symmetry only mildly in our simulations; the distributions 
${\cal P}_{n_cn_b}(w)$ depend on the coordination number 
$n_c+n_b$ rather than on $n_c$ or $n_b$ independently 
(Fig.~\ref{fig.abovebelow1}). 
This further refines the analysis of the relation between 
packing geometry and force network statistics in the bulk of 
a packing; at the boundary, it is sufficient to consider 
only the number of contacts from above ($n_c$).

\section{Weight and force distributions in
 the $q$-model: the role of connectivity.}\label{sec.qmodel1}

In this section, we investigate to what extent the results obtained
for the Hertzian sphere packings can be understood within the context of
the $q$-model and its generalizations. In the standard version of
the model, the particles are positioned on a regular lattice, and the
particle weights are stochastically transmitted to the neighbors in
the layer below \cite{qm}. The weight on a particle $i$ splits up into $n_c$
fractions $q_{ij}$, and the total weight exerted on a particle $j$ in
the layer below then becomes
\begin{equation}\label{qmodeldef}
W_j = mg + \sum_i q_{ij}W_i,
\end{equation}
where the term $mg$ can be neglected at large depth. The fractions
$q_{ij}$ obey the constraint
\begin{equation}\label{qconstraint}
\sum_j q_{ij}=1,
\end{equation}
which assures mechanical equilibrium in the vertical direction. They
can in principle also be deduced for more realistic packings: from
definition (\ref{defweight}), one finds $q_{ij}= \left (
\vec{F}_{ij}\right)_z/W_i$.

The simple form of the $q$-model has allowed for a number of exact
results of which the most important is the solution for the uniform
$q$-distribution. This uniform $q$-distribution assigns an equal
probability to each set of $\{q_{ij}\}$ that obeys
Eq.~(\ref{qconstraint}), and serves as a generic case. 
The rescaled weights $w$ then become distributed as \cite{qm}
\begin{equation}\label{qpw}
{\cal P}_{n_c}(w) = c\,\, w^{n_c-1} \, e^{-n_c\,w},
\end{equation}
where $n_c$ is fixed for a given lattice, and $c$ is a normalization
constant. Note that these solutions have the same qualitative behavior
as those found in our molecular dynamics simulations: for small
weights ${\cal P}_{n_c}(w) \propto w^{n_c-1}$, and the probability for
large weights decays exponentially.

The $q$-model is thus an effective minimal model for the weights $W$. 
It is clear that the product of $q_{ij}$ and $W_i$ has a natural
interpretion as the vertical component of $\vec{F}_{ij}$. Since these
interparticle forces are more fundamental than the weights, we
investigate the statistics of the quantity $qW$ in Sec.~\ref{sec.qw}; 
this will shed new light on the discrepancy for small forces 
between the $q$-model and experimental data. 
In the light of our finding that the contact geometry and in
particular $n_c$ plays a crucial role, the standard $q$-model is
clearly limited since it fixes $n_c$. In Sec.~\ref{sec.qrandcon} we
therefore extend the $q$-model to have randomness in its {\em
connectivity} (i.e. to allow for a range of $n_c$'s), and find that,
as expected, the ${\cal P}(w)$ can be manipulated by changes in the 
connectivity.

\subsection{Distribution of interparticle forces: $P(qw)$}\label{sec.qw}
A direct comparison of Eqs.~(\ref{defweight}) and~(\ref{qmodeldef})
shows that the product $q_{ij}w_i$ has a natural interpretation as the
vertical component of $\vec{f}_{ij}$. Since the interparticle forces
are more important than the weights, it is interesting to investigate
the statistical properties of the bond quantity $qw$. To obtain the
distribution $P(qw)$, let us start with the transformation from
$P(qw)$ to ${\cal P}_{n_c}(w)$:

\begin{eqnarray}
{\cal P}_{n_c}(w)&=&\int_0^\infty d(qw)_1 P\left(\!
\begin{array}{ll} \displaystyle (qw)_1 \end{array} \!\right) \cdots \nonumber \\
&&\times \int_0^\infty d(qw)_{n_c} P\left(\!
\begin{array}{ll}
\displaystyle (qw)_{n_c} \end{array} \!\right)  \nonumber \\
&&\times\, \delta\left( w- \sum_{i=1}^{n_c} (qw)_i\right).
\end{eqnarray}
Here we assumed that the $(qw)_i$ are uncorrelated, which is
valid for the uniform $q$-distribution \cite{jacqcorr}. For
the corresponding Laplace transforms, denoted by
$\tilde{P}(s)$ and $\tilde{\cal P}_{n_c}(s)$ respectively,
this relation becomes
\begin{equation}\label{connectionwf}
\tilde{{\cal P}}_{n_c}(s) = \left(\tilde{P}(s)\right)^{n_c}~.
\end{equation}
Since the Laplace transform of Eq.~(\ref{qpw}) is of the form 
$1/(1+s)^{n_c}$, the distribution of $qw$ reads:
\begin{equation}\label{qpf}
\tilde{P}(s) = \frac{1}{1+s} \quad\Rightarrow \quad P(qw)=
e^{-qw}.
\end{equation}
We thus find (for the uniform $q$-distribution) that $P(qw)$ is a pure
exponential, independent of the number of contacts $n_c$. Again, this
is very similar to the results for our Hertzian sphere packings: the
distribution of ``interparticle forces'' $P(qw)$ is finite for small
forces, whereas the distribution of weights depends on $n_c$ as given
by Eq.~(\ref{pncscaling}). Moreover, this resolves the discrepancy for
small forces mentioned in the introduction: the $q$-model predicts a 
vanishing probability densitity for small weights, {\em but not} for
small forces.


\subsection{Including geometry effects}\label{sec.qrandcon}
\begin{figure}[t]
\includegraphics[width=3.5cm]{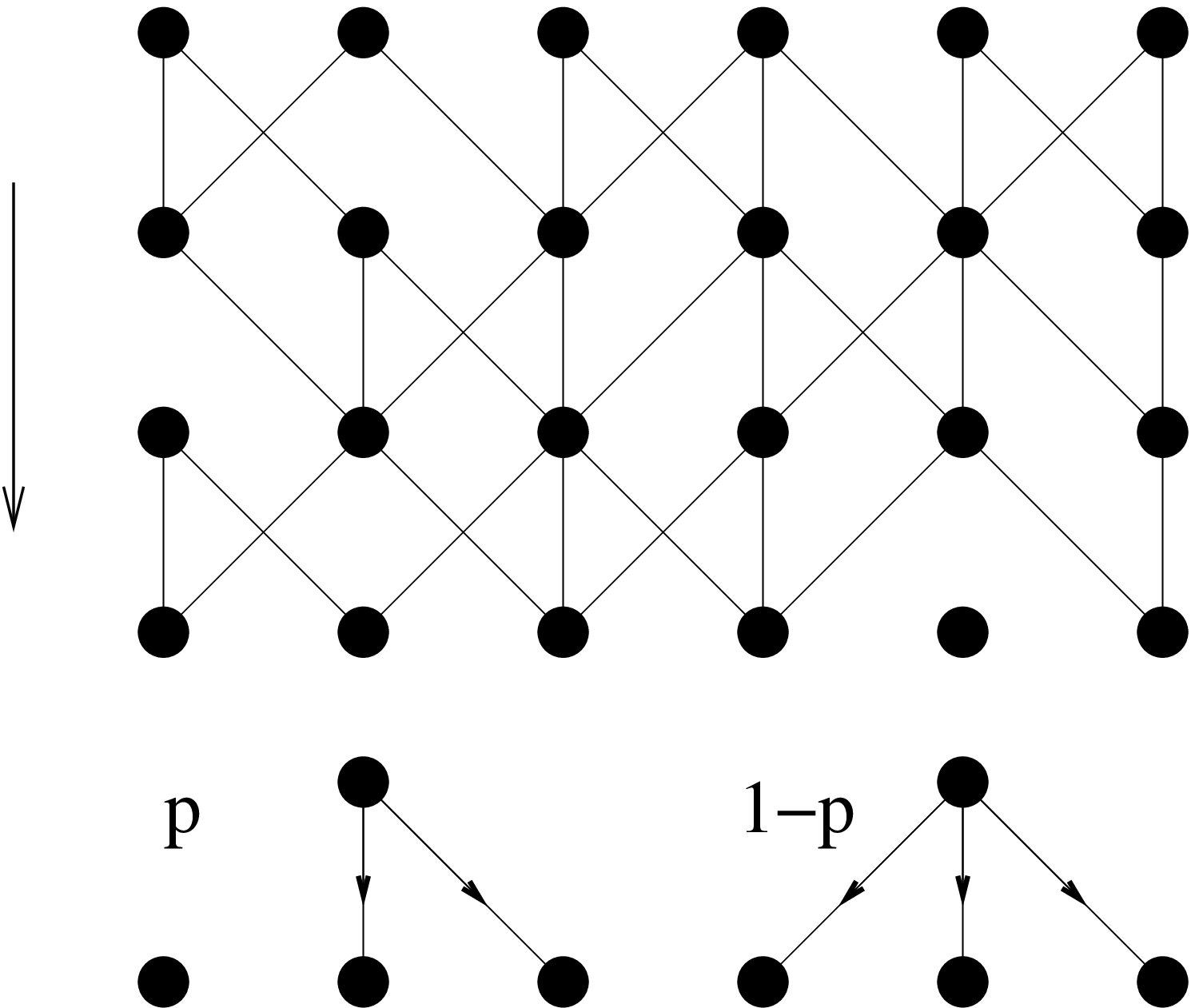}
\includegraphics[width=5.0cm]{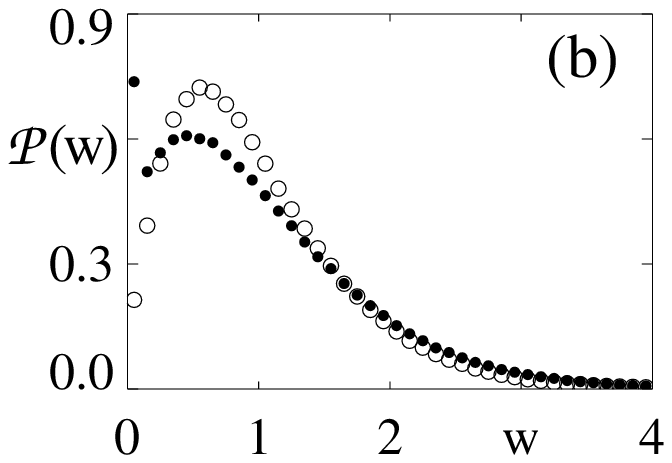}
\caption{The $q$-model with a random connectivity: 
{\em (a)} with a probability $p$ we cut one of the three bonds; 
{\em (b)} bottom effect in the $q$-model with random connectivity. 
In the bulk $p=0.3$ (open circles) and at the bottom $p=0.9$ (dots); 
this corresponds to
$\{\rho_0,\rho_1,\rho_2,\rho_3\}=\{0.00,0.03,0.24,0.73\}$ and
$\{0.03,0.19,0.44,0.34\}$ respectively.} \label{fig.qrandcon}
\end{figure}

From Sec.~\ref{sec.pw}, it is clear that the weight distribution
${\cal P}(w)$ in Hertzian sphere packings is very sensitive to the
local packing geometry. Since the $q$-model is defined on a regular
lattice, with fixed connectivity, it can not capture the behavior of
${\cal P}(w)$ in disordered packings with fluctuating $n_c$. This
extra degree of disorder can be included, for example, by ``cutting''
some of the bonds of the regular lattice. We illustrate this with the
2-dimensional square lattice depicted in
Fig.~\ref{fig.qrandcon}a. For each site, the weight is transmitted
downwards through either 2 or 3 bonds with probabilities $p$ and $1-p$
respectively; in the former case we randomly cut one of the available
bonds and generate the two remaining $q_{ij}$ according to a uniform
distribution satisfying Eq.~(\ref{qconstraint}). This generates
particles with $n_c=0,1,2\;{\rm and}\; 3$, since all bonds {\em
arriving} at a site have a probability of $p/3$ to be missing. 
For simplicity, we introduced the disorder in $n_c$ by means of one 
parameter $p$ only; 
as a consequence, we can only obtain a limited set of $\{\rho_{n_c}\}$.

With this model, we have tried to mimic the bulk-bottom behavior of 
${\cal P}(w)$ that was observed in the amorphous packings 
(Fig.~\ref{fig.pwamorph}a). 
In the bulk layers we took out bonds with probability $p=0.3$, and 
for the bottom layer we took $p=0.9$; the result is shown in 
Fig.~\ref{fig.qrandcon}b. Indeed, the change in the fractions 
$\rho_{n_c}$ is sufficient to reproduce a transition of ${\cal P}(w)$ 
reminiscent of what has been observed in our Hertzian sphere packings 
(compare with Fig.~\ref{fig.pwamorph}a).

\subsection{Conclusions for the $q$-model}

Although it is known that the $q$-model does not properly 
describe the spatial structure of the force-network \cite{greens}, 
it remains a very instructive theoretical framework for the {\em statistics} 
of force fluctuations. While in the standard case the
disorder in the system is represented by the stochastic fractions
$q_{ij}$ only, we have shown that when also the connectedness is chosen
to be random, the model displays most features of realistic packings.

Let us conclude this section by mentioning that the idea to leave out
some of the bonds of a regular lattice is not new \cite{cutbonds}. In
these studies, however, bonds were cut in a particular manner to build
up directed force chains. We have shown that such long-ranged
structures are not important for the behavior of ${\cal P}(w)$, since
they only depend on the {\em local} packing geometry.

\section{Top-down relaxation of fluctuations}\label{sec.qmodel2}

So far, the discussion has been limited to situations well below the 
top surface of the packings. The data of the Hertzian sphere 
simulations were taken at least 15 layers below the top surface and 
the results of the $q$-model (presented in the previous section) 
all correspond to the limit of large depths. In this section 
we investigate the top-down relaxation of the force and 
weight distributions. 
At the top surface of the Hertzian sphere packings, 
there are only weight fluctuations due to grain polydispersity.
The question we address is how fast the force and weight fluctuations 
build up towards a bulk distribution, as a function of depth. 

These results can then be compared to the relaxation in the $q$-model. 
Interpreting the downward direction as time, this corresponds 
to transient behavior towards the ``stationary'' solutions given in 
Eqs.~(\ref{qpw}) and~(\ref{qpf}). This top-down relaxation of 
fluctuations forms an additional test to qualify various theoretical 
models, very much like the Green's function measuring the response 
to a localized load on the top-surface \cite{greens}. 
In our case, we start from spatially (nearly) homogeneous conditions 
in the top layer and see how fluctuations build up.

\subsection{Top-down relaxation in Hertzian sphere packings}
\begin{figure}[t]
\includegraphics[width=8.5cm]{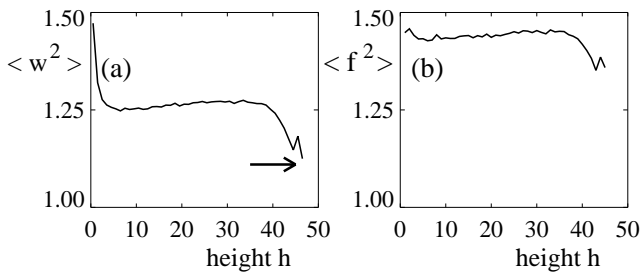}
\caption{The second moments {\em (a)} $\langle w^2\rangle$ and 
{\em (b)} $\langle f^2 \rangle$ as a function of height $h$ in 
simulations of Hertzian sphere packings. The arrow indicates the 
location of the top surface, around $h=46$. Both for the forces and the 
weights one finds a fast top-down relaxation of the moments.}
\label{fig.hertzrelax}
\end{figure}

A good way to quantify changes in ${\cal P}(w)$ and $P(f)$ is to study 
their second moments $\langle w^2\rangle$ and $\langle f^2 \rangle$. 
For a distribution of zero width these second moments are unity, 
and they increase as the fluctuations become larger. 
In Fig.~\ref{fig.hertzrelax} we show the second moments as a 
function of the height $h$, which is defined as the distance from 
the bottom boundary. Since the packings are strongly disordered, 
the precise location of the top surface will be slightly different 
for each realization; it turns out to be located around $h=46$. 

Let us first consider the broadening of the weight distribution 
shown in Fig.~\ref{fig.hertzrelax}a. As already mentioned above, 
the weight fluctuations at the top surface are entirely due to 
polydispersity of the grains. 
Using a flat distribution between $0.4 < r< 0.6$ this corresponds to 
$\langle w^2 \rangle \approx 1.11$, which is consistent with our 
simulation data. The second moment approaches its bulk value 
already at a depth of approximately 10 particle diameters. 
The figure also shows the sharp transition of ${\cal P}(w)$ at the 
bottom boundary. The second moments of $P(f)$ are shown in 
Fig.~\ref{fig.hertzrelax}b. One again observes a relaxation 
over approximately 10 layers, towards a bulk value; 
$P(f)$ does not change significantly near the bottom boundary. 
Note that both the force and weight distributions become 
slightly narrower as the depth increases below heigths of the 
order of $30$. This may be attributed to an increase in particle 
deformations \cite{makse}.

We thus find that the typical length scale for force and weight 
fluctuations to saturate is approximately 10 particle diameters. 
This provides another important criterion to distinguish between 
different theoretical models.

\subsection{Top-down relaxation in the $q$-model}
The top-down relaxation is well understood for the $q$-model without 
the so-called {\em injection term}, i.e. $mg=0$ in Eq.~(\ref{qmodeldef}) 
\cite{lew,jacqcorr}. 
Before extending these results to the $q$-model with injection, 
we briefly recapitulate the results of the $q$-model without the 
injection term $mg$. This version of the model can be interpreted 
as a packing of weightless particles, supporting a homogeneously
applied force. To distinguish between the $q$-model without injection 
from the model with injection, we denote the weight distributions 
at depth $t$ by ${\cal R}^{(t)}(w)$ (without injection) and by 
${\cal P}^{(t)}(w)$ (with injection). 

For the uniform $q$-distribution, it has been shown that
\cite{jacqcorr}
\begin{equation}\label{pwrelax}
{\cal R}^{(t)}(w)\, \simeq \,{\cal P}(w) +
\left(\frac{1}{\sqrt{t}}\right)^{d-1}{\cal F}(w) \quad\quad
{\rm for}\quad t \rightarrow \infty,
\end{equation}
where $d$ is the dimensionality of the packing. The stationary
solution ${\cal P}(w)$ is given by Eq.~(\ref{qpw}) and ${\cal
F}(w)$ is the shape of a typical deviation. It is
clear that all second and higher order moments $\langle w^k
\rangle$ approach their asymptotic values according to the
same power-law. This slow relaxation towards ${\cal P}(w)$ is
caused by {\em diffusion of correlations}, which takes place
in the ($d-1$)-dimensional correlation space \cite{footprint}.

\begin{figure}[t]
\includegraphics[width=8.5cm]{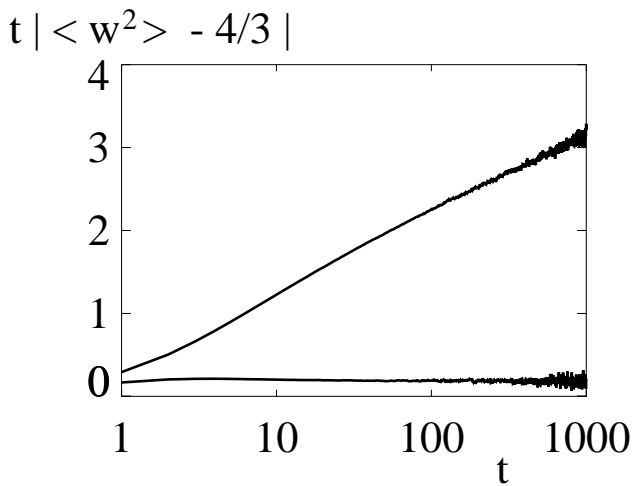}
\caption{Relaxation of the second moments 
with injection (climbing line) and without injection (flat line) 
towards their asymptotic values $4/3$ in the 3D $q$-model. 
Since we plot $t\times|\langle w^2 \rangle^{(t)} - 4/3|$ along the 
vertical axis, the climbing straight line confirms the $\log(t)/t$ 
relaxation for the $q$-model with injection. 
Without injection the relaxation is simply $1/t$.}
\label{fig.logrelax}
\end{figure}

Let us now investigate how the injection term $mg$ affects the
top-down relaxation. We first note that the recursive relation
for the weights, Eq.~(\ref{qmodeldef}), is a linear equation.
The $q$-model with injection can therefore be interpreted as a
superposition of $q$-models without injection, with
differently positioned initial layers. Although it is not a
priori clear how this superposition property is reflected in
the weight distributions ${\cal P}^{(t)}(w)$ (with injection)
and ${\cal R}^{(t)}(w)$ (without injection), we propose the
following approximate mapping:
\begin{equation}\label{mapping1}
{\cal P}^{(t)}(w)= \, \frac{1}{t+1}\sum_{t'=0}^t {\cal
R}^{(t')}(w).
\end{equation}
If we combine this with the exact result of
Eq.~(\ref{pwrelax}), we obtain the following relaxation as $t
\rightarrow \infty$:
\begin{eqnarray}\label{mapping2}
{\cal P}^{(t)}(w)- {\cal P}(w) &\propto& {\cal F}(w) \,\,
\, \frac{1}{t+1}\sum_{t'\neq 0}^t \left(\frac{1}{\sqrt{t'}}\right)^{d-1} \nonumber \\
\nonumber \\
&\propto& {\cal F}(w) \,\left\{ \begin{array}{ll}
\displaystyle \frac{1}{\sqrt{t}} & \mbox{$d=2$~,} \\ \\
\displaystyle \frac{\log(t)}{t} & \mbox{$d=3$~,} \\ \\
\displaystyle \frac{1}{t} & \mbox{$d\geq4$~.} \end{array}
\right.
\end{eqnarray}
This relaxation behavior is indeed observed in our numerical
simulations with $d=2$ and $d=3$, using a uniform 
$q$-distribution. In Fig.~\ref{fig.logrelax}, we show the
results for an fcc packing ($d=3$). We plot
$t\times|\langle w^2 \rangle^{(t)} - 4/3|$ as function of depth $t$, 
where $\langle w^2 \rangle^{(\infty)}=4/3$. 
The climbing straight line on the lin-log plot confirms
the remarkable $\log(t)/t$ relaxation. We also plot the same
data for the $q$-model without injection; this curve becomes
flat in agreement with Eq.~(\ref{pwrelax}).

Although the mapping of Eq.~(\ref{mapping1}) is definitely not
exact, it apparently captures the main physics of the
relaxation process. This can be understood as follows. There
are two slow processes involved: {\em (i)} the increasing
number of layers reduces the contribution of each layer of
``injected'' weights effectively as $1/t$; {\em (ii)} each
layer of injected weights relaxes as $(1/\sqrt{t})^{(d-1)}$
individually. Naturally, the total relaxation is dominated by
the slower of these two processes. In the special case of
$d=3$ both powers are $1/t$, leading to a logarithmic
correction. 
Finally note that since the downward 
$q$-values are statistically independent from the weights, the
``force'' fluctuations simply follow from $\langle (qw)^2
\rangle=\langle q^2 \rangle \langle w^2 \rangle$, and thus
display the same relaxation as the weights fluctuations.

\subsection{Conclusions concerning top-down relaxation}
We have studied the top-down relaxation of the second moments 
$\langle w^2 \rangle$ and $\langle f^2 \rangle$, which quantifies 
how `fast' the weight and force distributions approach their 
bulk shapes. The $q$-model predicts a power-law relaxation with a 
logarithmic correction for 3D packings, Eq.~(\ref{mapping2}). 
However, we find no evidence for such a slow relaxation in our simulations 
of Hertzian spheres, which indicate that a bulk distribution is 
reached after approximately 10 layers of particles 
(Fig.~\ref{fig.hertzrelax}). In the $q$-model with injection, 
for example, the second moment after $10$ layers still differs 
around $20\%$ from its asymptotic value. 

Let us provide two possible explanations why the $q$-model fails 
to describe this relaxation process. A first problem of the model is 
that it assumes some fixed $q$-distribution $\eta(q)$: we have seen 
that the $q$'s can in principle be derived from the forces as 
$q_{ij}=\left(\vec{F}_{ij}\right)_z/W_i$, so a relaxation in 
$P(f)$ and ${\cal P}(w)$ should result into a relaxation of 
$\eta(q)$ itself. This clearly shows the difficulty of encoding the 
force behavior into a stochastic variable $q$ in a self-consistent manner. 
Another problem of the model is that it assumes a top-down propagation 
of forces. The up-down symmetry is therefore broken explicitly in the 
$q$-model, whereas in our Hertzian sphere packings we find only a very 
weak symmetry breaking. 
In principle, force networks are defined by the equations of 
mechanical equilibrium, which generically are {\em underdetermined} 
\cite{grestcoord,force-ensemble} and hence can not be solved by an iterative 
(top-down) procedure. 
Instead, one has to solve this set of coupled equations ``simultaneously'' 
for all particles in the system, and except for the (small) $mg$ term, 
there is a natural up-down symmetry in this system. 
The absence of this up-down symmetry in the $q$-model could of 
course strongly affect the top-down relaxation.

\section{Discussion}

We have shown that in order to understand the statistics of force 
networks, it is crucial to distinguish between forces and weights. 
We have found in our simulations that the 
force distribution $P(f)$ is very robust, in the sense that its 
shape does not depend on details of packing geometry. 
The weight distribution ${\cal P}(w)$, on the other hand, 
is very sensitive for the local packing geometry. 
We have demonstrated that a decomposition according to the number of 
contacts that press on a particle from above, $n_c$, is sufficient 
to understand this geometry dependence. 
Reinterpreting experiments on strongly deformed rubber particles 
\cite{rubber} within this framework, 
we find strong evidence that $P(f)$ essentially remains unaffected 
even by very large particle deformations. 
To further test our framework experimentally, one can manipulate 
the number of contacts at the boundary by placing a layer of 
relatively small or large beads at the bottom. 
For small beads, the fractions $\rho_0$ and $\rho_1$ will be enhanced, 
leading to a large ${\cal P}(w)$ for small $w$, 
and a slow exponential decay for large $w$. Relatively large 
bottom beads should lead to a ${\cal P}(w)$ that is strongly peaked.

The present work provokes a number of questions. 
First, we observe that most of our simulation results, like the 
shapes of $P'(f_z)$ and ${\cal P}(w)$, 
can be understood in terms of {\em local} packing geometry only. 
This suggests that long-range correlations are not dominant, 
at least not for the `one point' force, weight and angle 
probability distributions. 
We therefore question whether the behavior of $P(f)$ observed 
at the jamming transition \cite{liuletter,grestjam} reflects a 
long-range structural change of the force network. 
In particular, we expect that the role of `force chains' can only 
be understood from two or more-point correlation functions, 
and not from $P(f)$ only. 

A related problem is that the $q$-model fails to describe problems 
that involve spatial structure of the force network. 
Although the model is able to capture many features of force 
and weight statistics (Sec.~\ref{sec.qmodel1}), 
it does not produce the top-down relaxation of 
${\cal P}(w)$ that is observed in the more realistic Hertzian packings. 
Alongside with the incorrect prediction of the response 
function \cite{greens}, this indicates that spatial dependence is not 
correctly incorporated within the $q$-model. 
This may be due to the fact that, in general, recursive models do 
not acknowledge the structure of the equations describing 
mechanical equilibrium. These equations are typically underdetermined 
\cite{grestcoord} and cannot be solved in a recursive manner. 
In a recent paper \cite{force-ensemble}, we therefore propose a 
different theoretical approach, in which we start from the equations 
of mechanical stability and exploit the undetermined degrees of freedom. 

Another important issue for future study is clearly the role of 
friction and dimensionality. Our numerical study has been done in two
dimensions with frictionless spheres; however, recent studies indicate
\cite{makse} that the coordination number for 3D packings with
friction is similar to those of 2D frictionless
packings. Qualitatively, the picture we have advanced is therefore
expected to capture the realistic case of three dimensions with
friction, because our phase space arguments are independent of
dimension.

{\bf Acknowledgements}
We are very grateful to Nathan Mueggenburg and Heinrich Jaeger for 
providing some of their experimental data and for the open exchange 
of ideas. We also thank Martin Howard, Hans van Leeuwen and 
Carlo Beenakker for numerous illuminating discussions. 
JHS and ES gratefully acknowledge support from the 
physics foundation FOM, and MvH support from the science foundation 
NWO through a VIDI grant.

\appendix
\section{Logarithmic divergence of $P'(f_z)$}\label{app.logdiv}
In Sec.~\ref{subsec.orientation}, we encounter the following
integral:
\begin{eqnarray}\label{a1}
P'(f_z)&=&\int_0^{\pi}d\varphi \,\frac{1}{\pi}\, \int_0^\infty
df \,P(f)\,
\delta\left(f_z-f\sin(\varphi)\right) \nonumber \\
&=&\int_0^\infty df
\,P(f)\,\int_0^{\pi/2}d\varphi\,\frac{2}{\pi}\,
\frac{1}{f}\,\delta\left(\frac{f_z}{f}- \sin(\varphi)\right) \nonumber \\
&=& \frac{2}{\pi}\int_{f_z}^\infty df
\,\frac{1}{\sqrt{f^2-f_z^2}}\, P(f)~.
\end{eqnarray}
The function $P(f)$ represents the probability density
function of $f=|\vec{f}|$, which we can assume to be regular
on the entire interval (see Fig.~\ref{fig.pfamorph}). The
behavior for small $f_z$ is not trivial, since the integrand
diverges at the lower bound of the integration interval. For
each non-zero $f_z$ this does not lead to a singularity, since
\begin{eqnarray}\label{a2}
P'(f_z)&=& \frac{2}{\pi}\int_{f_z}^\infty \frac{df}{f_z} \,
\frac{P(f)}{\sqrt{\left(f/f_z\right)^2-1}} \nonumber \\
&=& \frac{2}{\pi} \int_1^\infty du
\frac{P(uf_z)}{\sqrt{u^2-1}}~.
\end{eqnarray}
The integral over $1/\sqrt{u^2-1}$ is convergent for
$u\rightarrow1$ and the function $P(uf_z)$ falls of fast
enough as $(uf_z) \rightarrow \infty$. For $f_z=0$, however,
the integral diverges as $u\rightarrow \infty$. To obtain the
asymptotic behavior we rewrite the integral as
\begin{eqnarray}
P'(f_z) &=& \frac{2}{\pi}\int_1^\infty du \frac{P(uf_z)}{u} \nonumber \\
&+&\,\frac{2}{\pi}\int_1^\infty du
P(uf_z)\left(\frac{1}{\sqrt{u^2-1}}- \frac{1}{u}\right)~.
\end{eqnarray}
The second term is convergent since the term between brackets
behaves as $1/u^3$ in the limit $u\rightarrow \infty$. We thus
find that
\begin{eqnarray}
P'(f_z \rightarrow 0) &\simeq& \frac{2}{\pi}\int_{f_z}^\infty
df \,
\frac{P(f)}{f}  + {\cal O}(1) \nonumber \\
&\simeq& -\frac{2}{\pi}\, P(0) \, \ln(f_z) + {\cal O}(1)~.
\end{eqnarray}

\section{Relation between tails of $P'(f_z)$ and ${\cal P}_{n_c}(w)$}
\label{app.tails}
In this appendix we derive the large weight behavior of ${\cal
P}_{n_c}(w)$ from the tail of $P'(F_z)$, assuming that the
various $\left(\vec{F_i}\right)_z$ in Eq.~(\ref{trafo}) are
uncorrelated. We consider decays both faster and slower than
exponential, of the form

\begin{equation}\label{c1}
P'(F_z) \propto e^{-\alpha F_z^\beta/\langle F_z \rangle^{\beta}} 
\quad {\rm for}\quad F_z
\rightarrow \infty~.
\end{equation}
We show that, after rescaling $\langle w \rangle$ to unity,
this leads to
\begin{equation}\label{c2}
{\cal P}_{n_c}(w)\propto e^{-\gamma w^\beta}~,
\end{equation}
with
\begin{eqnarray}\label{c3}
\gamma &=& \,\left\{ \begin{array}{ll}
\displaystyle \alpha \, \,n_c & \mbox{, $\beta\geq1$} \\ \\
\displaystyle \alpha \, \,n_c^{\beta}
& \mbox{, $\beta\leq1$~.}
\end{array} \right.
\end{eqnarray}
This means that the tail of the weight distribution is of the
same nature as that of the forces, but with a different
prefactor $\gamma$. The tails get steeper for increasing
$n_c$, since the reduced probability for small $w$ (due to a
lack of phase space) must be compensated to keep $\langle w
\rangle=1$.

The above results are obtained as follows. Rescaling all
forces in Eq.~(\ref{trafo}) as $x_i=\left(F_z\right)_i/W$, one
obtains the probability for large weights

\begin{equation}\label{c4}
{\cal P}_{n_c}(W) \propto W^{n_c-1}\int_{\cal S} dx_1\cdots
dx_{n_c} \, e^{-\frac{\alpha}{\langle F_z\rangle^\beta} 
W^{\beta} \left(x_1^\beta + \cdots +
x_{n_c}^\beta \right)}~,
\end{equation}
where ${\cal S}$ denotes the hyperplane $1-\sum_i x_i$ with 
all $x_i\geq0$.

For $\beta>1$, the probability density on ${\cal S}$ has a
maximum at $x_i=1/n_c$, which becomes sharply peaked for
increasing $W$. Physically, this means that the dominant
contribution for large weights will come from all $F_z$ being
equal, namely $W/n_c$. Approximating the integrand by a
Gaussian around its maximum value, we find that the ``width''
decreases as a power of $W$ only, namely
$1/W^{(n_c-1)\beta/2}$. Hence the leading behavior for large
$W$ is given by the maximum value of the integrand, i.e.
$e^{-\frac{\alpha}{\langle F_z \rangle^\beta} W^\beta/(n_c)^{\beta-1}}$.

For $\beta<1$, the probability density has a minimum at
$x_i=1/n_c$, and the dominant contribution now comes from
$x_i=1$ and $x_{j\neq i}=0$. This means that typically only
one of the forces accounts for the whole weight. The part of
the integral around $x_i=1$ can be approximated by

\begin{equation}
e^{-\frac{\alpha}{\langle F_z \rangle^\beta} W^\beta} \int_{{\cal S}_\epsilon} dx_1 \cdots
dx_{n_c} e^{-\frac{\alpha}{\langle F_z \rangle^\beta} 
W^\beta \sum_{j\neq i}x_j^\beta}~,
\end{equation}
where ${\cal S}_\epsilon$ denotes the part of ${\cal S}$ for
which $1-x_i\leq \epsilon$. This approximation becomes exact
for $W\rightarrow\infty$ as long as $W^\beta \epsilon \ll 1$;
we take $\epsilon=1/W^{1-\delta}$ with $0< \delta <1-\beta$.
Working out the integration over ${\cal S}_\epsilon$, one
finds

\begin{equation}
\frac{e^{-\frac{\alpha}{\langle F_z \rangle^\beta} 
W^\beta}}{W^{n_c-1}}\left( \int_0^\infty dy
\,e^{-\frac{\alpha}{\langle F_z \rangle^\beta} y^\beta} \right)^{n_c-1}~,
\end{equation}
as $W\rightarrow \infty$. The part of the integral outside the
areas ${\cal S}_\epsilon$ is smaller than $W^{n_c-1}e^{-
\frac{\alpha}{\langle F_z \rangle^\beta}
W^\beta \left(1+W^\delta\right)}$ and can thus be neglected.
So also for $\beta<1$, the leading behavior for large $W$ is
simply given by the maximum value, i.e. $e^{-\frac{\alpha}{\langle 
F_z \rangle^\beta} W^\beta}$.

As mentioned in Sec.~\ref{sec.pw}, the ${\cal P}_{n_c}(W)$
obtained by Eq.~(\ref{trafo}) are not properly normalized,
since $\langle W \rangle=\langle f_z\rangle n_c$. If we
rescale the average weight to unity, we obtain the results of
Eqs.~(\ref{c2}) and~(\ref{c3}).


\begin{thebibliography}{99}
\bibitem{footnc} When determining the value of $n_c$, we explicitly exclude the
intralayer bottom contacts; for the strongly polydisperse packings, 
these correspond to angles that deviate less than $13^{\circ}$ from 
the horizontal. For our estimates of the bulk $\rho_{n_c}$'s we do not 
exclude any forces, i.e., the ``cutoff angle'' is strictly 
$0^{\circ}$. Since the bulk is isotropic, a cutoff angle of 
$13^{\circ}$ leads to a change of the order of 15\%, far too little to 
explain the change of $\rho_1$ by a factor of more than 4. For 
simplicity we therefore have kept a ``cutoff angle'' of $0^{\circ}$ in 
the bulk.
\bibitem{foottails} To determine the precise value of ${\cal P}_1(0)$ 
is a very subtle problem. Since the total weight distribtution 
${\cal P}(w)$ is normalized such that $\langle w \rangle=1$, the 
average weight of particles with $n_c=1$ will be smaller than unity 
and ${\cal P}_1(w)$ should be rescaled with a factor $\langle n_c\rangle$. 
Incorporating this in a self-consistent manner, we obtained very 
similar $\langle n_c \rangle$ as those presented in 
Table~\ref{tabrubber}, where we took ${\cal P}_1(0)=0.5$ 
for simplicity.
\bibitem{footprint} This diffusion in correlation space should not be confused
with the diffusion in real space due to a localized force on
the top surface. In our case we have homogeneous initial
conditions as all weights in the top layer are equal.
\end{thebibliography}
\end{document}